# Study on the imaging ability of the 2D neutron detector based on MWPC

TIAN LiChao[1;2;3*], CHEN YuanBo[1;2], TANG Bin[1;2], ZHOU JianRong[1;2], QI HuiRong[1;2], LIU RongGuang[1;2], ZHANG Jian[1;2], YANG GuiAn[1;2], XU Hong[1;2], CHEN DongFeng[4] & SUN ZhiJia[1;2]

1 State Key Laboratory of Particle Detection and Electronics, Beijing 100049, China
2 Institute of High Energy Physics, Chinese Academy of Sciences, Beijing 100049, China
3 University of Chinese Academy of Sciences, Beijing 100049, China
4 China Institute of Atomic Energy, Beijing 102413, China

**Abstract**

A 2D neutron detector based on $^3$He convertor and MWPC with an active area of 200 mm × 200 mm has been successfully designed and fabricated. The detector has been tested with Am/Be neutron source and with collimated neutron beam with wavelength of λ = 1.37 Å. A best spatial resolution of 1.18 mm (FWHM) and good linearity were obtained. This is in good agreement with the theoretical calculations.

**Introduction**

Two-dimensional imaging neutron detectors are being increasingly used in neutron diffraction experiments as the developments of the spectrometers at the China Spallation Neutron Source (CSNS) and the China Advanced Research Reactor (CARR). Thermal neutron detectors based on $^3$He neutron convertor and multi-wire proportional chamber (MWPC) are widely used because of its advantages of ease large area building, high position resolution, high efficiency and low gamma sensitivity [1-3]. However most of the two dimensional position sensitive neutron detectors used in China are bought from foreign sources. Domestic research on imaging neutron detectors has not been earnestly done in the past. Consequently, a two-dimensional position sensitive multi-wire detector with an active area of 200 mm × 200 mm for the Multifunctional Reflectometry (MR) of CSNS has been developed at Institute of High Energy Physics, Chinese Academy of Sciences (IHEP). The detector has been tested with Am/Be neutron source in IHEP and reactor neutron beam in China Institute of Atomic Energy (CIAE). A good spatial resolution and excellent linearity are obtained. The test results and the analysis are reported in this paper.

**Detector description**

The neutron, a neutral particle, is absorbed by $^3$He atom in the detector and two energetic charged ions, a 191 keV triton and a 573 keV proton, are emitted in opposite directions as the reaction of (1). The ionized electrons by proton and triton drift under electric field and an avalanche takes place on the nearest anode wire. Inducted signals appeared on the neighboring readout strips will be recorded and the neutron position can be reconstructed though the center of gravity method. This is the basic principle of the neutron detection.

$$^3\text{He} + n \rightarrow {}^1\text{H} + {}^3\text{H} + 764\text{keV} \qquad (1)$$

The imaging quality usually depends on the signal to noise ratio (SNR) and the pixel size, which are accordant with the n/γ suppression ability and the position resolution for neutron imaging devices. To achieve the best position resolution, the key parameters of the MWPC have been optimized in the prototype research. Then, a two-dimensional spatial sensitive neutron detector constituting of the cathode-readout-anode-readout array, with an active area of 200 mm × 200 mm and a total gas depth of about 16 mm has designed and fabricated [4]. To meet the characteristics of high neutron detection efficiency and high position resolution, the MWPC is sealed in a high pressure chamber with 6 atm. $^3$He used as the neutron conversion medium and 2.5 atm. $C_3H_8$ as the stopping gas. The structure of the MWPC is as follows (also seeing Fig. 1).

- **Cathode plane:** Composed of aluminium foil with a thickness of 50 μm.
- **Anode plane:** Composed of gold-plated tungsten wires with a diameter of 15μm, the wire tension is 25 g, and the wire distance of 2 mm; all wires are connected together to get the deposited energy in the chamber and supply the trigger for data acquisition system (DAQ).
- **X readout plane:** Composed of gold-plated tungsten wires with a diameter of 50μm, the wire tension is 40 g, and the wire distance of 1 mm; the wires are parallel to the anode wires; 4 wires are connected together to form one readout strip.
- **Y readout plane:** Composed of a printed circuit board (PCB) covered with metal strips of 1.6 mm width; the strips are perpendicular to the anode wires; the metal strip distance is 2 mm and every two metal strips connected together to form one readout strip.

The readout strips are connected to the external electronic modules using ceramic to metal feedthroughs mounted on the rear panel of the container.

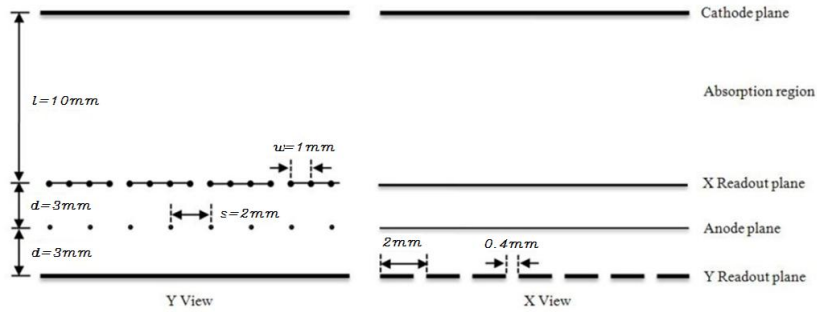

Figure 1 Schema for the planes of the MWPC

**Theoretical position resolution**

Position resolution in the direction perpendicular to anode wires is determined by the anode wire distance, which is 2 mm for this detector. The position resolution in the direction parallel to the anode wires can greatly improve through the center of gravity method.

The neutron position resolution is mainly determined by two aspects, which are the intrinsic resolution for electrons detection $\Delta x_e$ and the distribution of initial electron-ion pairs in the operation gas $\Delta x_{ion}$.

To know the intrinsic resolution for electrons detection, a prototype has been constructed and test by $^{55}$Fe with the flowing gas of Ar/CO$_2$ (90/10) prior to the neutron detector and a position resolution of 0.22 mm was obtained [4]. This can be considered to be the intrinsic resolution for electrons detection using the center of gravity readout method.

The moving range of the proton and triton created by the neutron conversion is the key factor for determining the position resolution of the detector. The protons and tritons are emitted uniformly in 4π and moving in opposite directions. The center of the ionization distribution has been displaced from the neutron absorption position because of the differential energy and moving range of proton and triton. The center of gravity for the initial ionization distribution by proton/triton pair is approximately 42% of the proton range [5]. For many injected neutrons, the centroids distribution can be described as a sphere. It is a uniform distribution projected in X or Y direction. In practice, range straggling of the triton and proton, and electronic noise, introduce a small broadening term, and the resulting distribution is shown by the dashed lines in Fig. 2. However the FWHM remains that of the uniform distribution [6]. The proton range in the operation gas of 6 atm. $^3$He + 2.5 atm. $C_3H_8$ can be obtained by the SRIM [7], as shown in Fig. 3. Thus, the positional uncertainty caused by the ion-range is about 84% of the proton range, which is 1.14 mm. The expected position resolution can be calculated by Eq. (2) and the result given as 1.16 mm.

$$\Delta x = \sqrt{\Delta x_e^2 + \Delta x_{ion}^2} = 1.16 mm \tag{2}$$

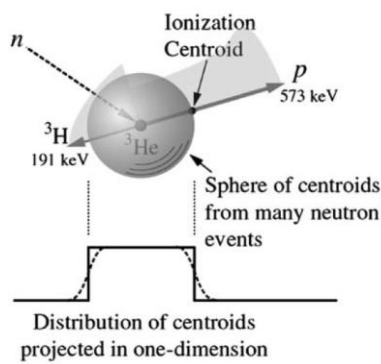 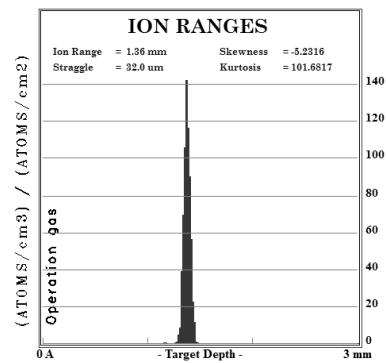

Figure 2 Thermal neutron conversions in $^3$He gas    Figure 3 Range of 573 keV protons in the operation gas

**Experimental Results**

**n/γ suppression ability**

The n/γ suppression of detector has been tested with the Am/Be neutron source in IHEP which is sufficient for uniform irradiation test, but not sufficiently powerful to generate collimated beams for high spatial resolution test. Figure 4 shows the energies deposited in the detector under different testing conditions. The spectrum (a) was tested with the $^{137}$Cs gamma ray source; the spectrum (b) was tested with an Am/Be source with a moderator and Pb in front of the window; the spectrum (c) was tested under the same condition with (b) but with 4 mm thick cadmium more in front of the window to absorb the thermal neutrons. Three spectrums had been normalized by the running time. With the appropriate threshold selected, the gamma rays could be discriminated from the neutrons easily through the comparison of the spectrum (a) with (b). [4]

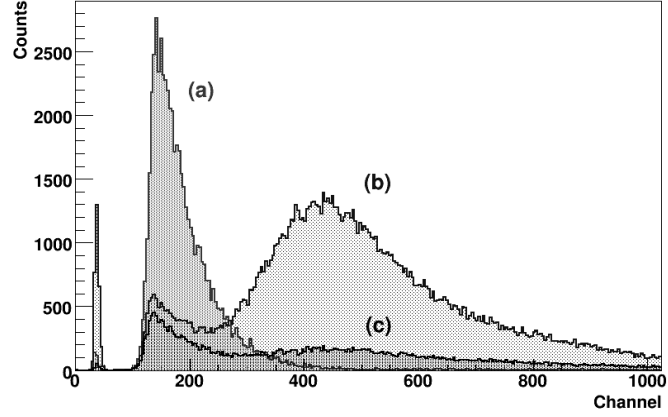

Figure 4 Energy spectra under different test conditions.

**Position resolution**

To measure the position resolution, the detector has been tested by a narrow neutron beam at the Residual Stress Diffractometer (RSD) [8,9] on CARR. The monochromator of RSD is flat copper (220) with wavelength 1.37 Å. The collimated neutron beam with a size of 5 mm $\times$ 40 mm, and a wavelength of 1.37 Å is incident at the center of the active area covered by a Cd-mask with three slits of different width in the space of 8 mm. The experimental equipment was set up as shown in Fig. 5.

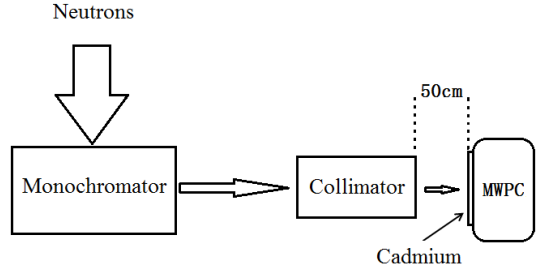

Figure 5 Schematic diagram of experimental setup

According to the central limit theorem, the position resolution function usually follows Gaussian distribution as given in Eq. (3). The neutron beam intensity (after going through the slit) on the surface of the detector, termed the beam intensity distribution, can be considered to be a uniform distribution as Eq. (4). Then, the distribution of reconstructed position spectrum, termed the experimental data distribution, can be described by a convolution of the beam intensity distribution and the detector resolution function, as seen in Eq. (5). [10]

$$f_g(x) = \frac{1}{\sqrt{2\pi}\sigma_0} e^{-\frac{(x-\mu)}{2\sigma_0^2}} \tag{3}$$

$$f_u(x) = \frac{1}{b-a} \tag{4}$$

$$f_{\exp}(x) = f_g(x) \otimes f_u(x) = \frac{1}{b-a}\left(F\left(\frac{b-x}{\sigma_0}\right) - F\left(\frac{a-x}{\sigma_0}\right)\right) \tag{5}$$

Where, $\mu$ and $\sigma_0$ are the mean value and the mean square deviation of the Gaussian distribution; $a$ and $b$ are the lower and upper limitation of the Uniform distribution; $F(x)$ is the cumulative distribution function of the Gaussian distribution.

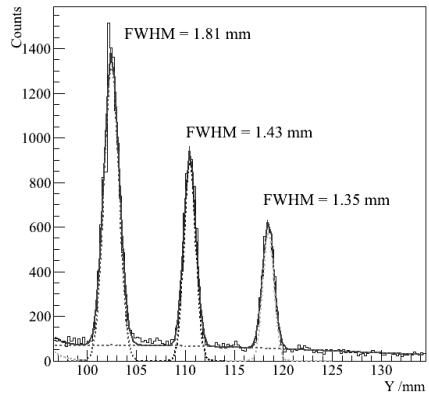

Figure 6 Position spectrum distribution

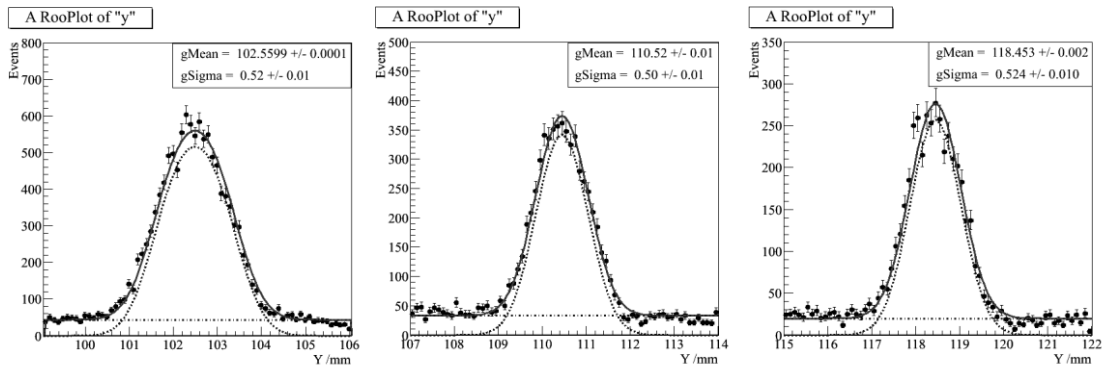

Figure 7 Fitted curve using the convolution of gauss and uniform function

Table 1 Position resolution obtained through different slits

| No. of slit | Slit width | Position spectrum (FWHM) | σ of the gauss function | Position resolution (FWHM) |
|---|---|---|---|---|
| 1 | 0.8mm | 1.35 mm | 0.52±0.01mm | 1.22±0.02 mm |
| 2 | 1.0mm | 1.43 mm | 0.50±0.01mm | 1.18±0.02 mm |
| 3 | 2.0mm | 1.81 mm | 0.52±0.01mm | 1.22±0.02 mm |

Fig. 6 shows the experimental data distribution taken at three positions through narrow slits on the center of active area. The peaks are fitted with the convolution of the gauss and uniform functions respectively and the results are shown in Fig. 7 and Table 1. A best position resolution (FWHM) of $0.50 \times 2.355 = 1.18$ mm is obtained, which is in good agreement with theoretical resolution.

**Holes distinguish ability**

To analyze the imaging ability, the detector was covered by another Cd-mask with a line of holes of different diameters and pitches in front of window and tested. The position spectrum projected in Y direction is shown in Fig. 8 and Table 2. The two holes in pitch of 2.0 mm with diameter of 1 mm can be clearly separated. While the two holes in pitch of 1.5 mm with diameter of 0.7 mm cannot be separated because the edges of the holes are in proximity to each other (0.8 mm). Concurrently, an excellent linearity was found using the seven holes (the indistinguishable two holes are not included), as seen in Fig. 9. The linearity can allow the image to be undistorted.

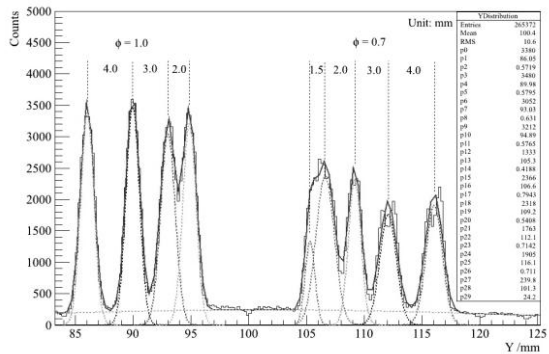

|  | Ø = 0.7 mm | Ø = 1.0 mm |
| --- | --- | --- |
| d = 1.5 mm | No | / |
| d = 2.0 mm | Yes | Half |
| d = 3.0mm | Yes | Yes |
| d = 4.0mm | Yes | Yes |

Figure 8 Position spectrum projected in Y direction

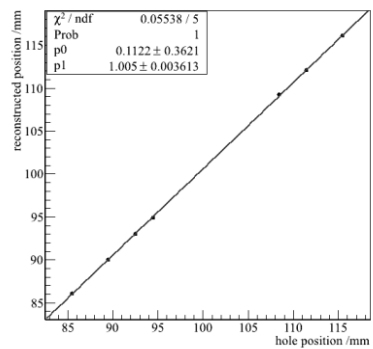

Figure 9 Linearity of the detector

**Conclusion**

A two-dimensional neutron imaging detector based on $^3$He convertor and MWPC with an active area of 200 mm × 200 mm has been successfully designed and fabricated. To obtain good position information, the center of gravity readout method is used. The detector is tested at CARR with a collimated neutron beam of λ = 1.37 Å. A best spatial resolution of 1.18 mm (FWHM) is obtained in the direction along the anode wires, which is in good agreement with the theoretical resolution. Good position linearity is also achieved.

**Acknowledgements**

The authors wish to acknowledge the many valuable contributions made to this project. The electronic engineering skills and data acquisition skills of ZHAO YuBin, ZHANG HongYu and ZHAO DongXu are gratefully acknowledged.

This work has been carried out at RSD of CARR in CIAE. We acknowledge, particularly, the grateful help of Prof. CHEN DongFeng, Prof. LIU YunTao, Dr. LI JunHong and Dr. YU ZhouXiang.